\documentclass[12pt,a4paper,final]{iopart}
\usepackage[utf8]{inputenc}
\usepackage{graphicx,color}
\usepackage{dcolumn}
\usepackage{bm}
\usepackage[breaklinks=true,colorlinks=true,linkcolor=blue,urlcolor=blue,citecolor=blue]{hyperref}
\usepackage{iopams}  
\usepackage{subfigure}
\usepackage[normalem]{ulem}
\usepackage{listings}
\usepackage{xcolor}

\usepackage{color}
\definecolor{mygreen}{rgb}{0,0.6,0}
\definecolor{mygray}{rgb}{0.5,0.5,0.5}
\definecolor{mymauve}{rgb}{0.58,0,0.82}

\begin{document}
\title{ClasSOMfier: A neural network for cluster analysis and detection of lattice defects}
\author{Javier F. Troncoso}
\address{Atomistic Simulation Centre, Queen's University Belfast, Belfast BT7 1NN, UK}%
\eads{\mailto{jfernandeztroncoso01@qub.ac.uk}}
\begin{abstract}
ClasSOMfier is a software package to classify atoms into a given number of disconnected groups (or clusters) and detect lattice defects, such as vacancies, interstitials, dislocations, voids and grain boundaries. Each cluster is formed by atoms whose atomic environment can be described by a common pattern. Unlike many methods available in the literature, where these patterns are given in advance and are associated with known lattice structures (i.e. fcc, bcc or hcp), this code implements a Kohonen network, which is based on unsupervised learning and where no information about the atomic environment has to be given in advance. ClasSOMfier accelerates the application of machine learning for cluster analysis by providing an efficient and fast code in Fortran with a user-friendly interface in Python.

\end{abstract}


\maketitle


\section{Introduction}\label{sec:level1} 
Some materials properties such as the mechanical strength or the thermal conductivity are determined by the lattice structure. Thus, the knowledge of the atomic structure is crucial in the design of materials for specific uses, and the design of defected lattices, including grain boundaries and point defects, has received significant attention recently. Just to give a concrete example: in PbTe, the thermal conductivity can be reduced by more than 30\% in the presence of grain boundaries and point defects \cite{javpabjor2019}. For that reason, it is important to study the dynamics and the interactions of these defects. The use of molecular dynamics (MD) simulations is an efficient strategy for this purpose, but they have to be supplemented by visualization tools to track the evolution of the position and size of all defects in the simulation box and their impact on the total energy of the system. From a computational point of view, the detection of these defects can be made by identifying patterns and crystal symmetries. For this purpose, several analysis methods have been proposed in the literature \cite{Stukowski2012}. The Common Neighbor Analysis (CNA) method and the Ackland–Jones analysis method are the most known methods. They study the environment of each atom and each atom is classified into a known ideal lattice structure (such as fcc, bcc or hcp) or an amorphous structure \cite{Honeycutt1987,Ackland2006}. In the CNA method, one atom is assigned to an ideal perfect lattice if its number of neighbors and the interatomic distances match the requirements of the lattice structures. In the Bond Angle Analysis (BAA) method, angles are used instead of distances.  The Ackland–Jones analysis method uses both distance and angular distributions. One of the advantages of these methods is that the number of ideal lattices is limited and the algorithm only needs information about the local environment of each atom, so they work relatively fast and are not limited to small samples. However, they are not able to characterize amorphous regions and certain information about the lattice structure has to be known in advance.

Other methods use specific properties of each atom, such as the potential energy,  to indicate if they are part of a perfect lattice. If one atom is part of a crystal defect, its energy is higher and it can be identified. Similarly, the identification of atoms of crystal defects can be achieved by using the Centrosymmetry parameter (CSP), a function of the interatomic distances among one atom and its first $N$ neighbors \cite{Kelchner1998}. The CSP and the potential energy per atom are useful quantities to characterize the local lattice disorder. However, in MD simulations, all atoms are far from their equilibrium positions, and the choice of the threshold energy or the cutoff distance that differentiates atoms that form the perfect lattice is not easy. 

Finally, geometric methods such as the Voronoi decomposition have also been proposed to identify the lattice structure \cite{Voronoi1908a}. In this method, one atom is enclosed by the Voronoi polyhedron formed by the first neighbors to determine the structural type. However, this method is highly sensitive to perturbations of the atomic positions, like those present in MD simulations. 

All of these methods can identify known lattice structures, but certain information about them has to be given in advance, and therefore they are not able to characterize amorphous regions. Thus, the development of improved methods can play an important role in the study and characterization of crystal structures, and the utilization of algorithms based on unsupervised learning can be an efficient tool for classification when no information about the structure is given. Machine learning methods are sophisticated tools that can extract information from large amounts of data from experiments or simulations \cite{Carleo2019}. These methods include techniques such as artificial neural networks (NN) \cite{Behler2016}, Gaussian processes \cite{Rasmussen2006}, and support vector machines \cite{Cristianini2000}. Machine learning is the study of computer algorithms with the ability to learn automatically through experience. Thus, a machine learning algorithm can be used for prediction or classification when it is trained to learn patterns in the data. This learning is usually classified as supervised learning, unsupervised learning, or reinforcement learning \cite{Carleo2019}. In supervised learning, the input and output variables are known, and the algorithm learns the mapping function that relates the inputs to the outputs. On the other hand, the output variables are not known in unsupervised learning and the data are classified according to common patterns in the input data. In reinforcement learning, the algorithm learns from the consequences of the actions. The goal of reinforcement learning is to learn a good strategy from experimental trials to perform future actions. Unsupervised learning algorithms can be further subdivided into clustering methods, dimensionality reduction schemes and generative models that produce new structures compatible with a known pattern \cite{Ceriotti2019}. Clustering methods classify input vectors into different groups according to common patterns and dimensionality reduction schemes produce a low-dimensional projection of high-dimensional data. In the present work, we use unsupervised learning for cluster analysis. 

The simplest unsupervised learning algorithm for cluster analysis is the k-means algorithm \cite{MacQueen1967}. This method starts with $k$ initial random points, known as seeds, and all input vectors are compared with all these seeds by means of the Euclidean distance and assigned to the cluster represented by the closest seed. After each step, the seeds are recalculated as the average position of the objects assigned to the cluster that they represent. However, this method is sensitive to initialization and depends on the choice of the initial seeds. More complex algorithms, such as the Fuzzy c-means, where each input vector can belong to more than one cluster,  and artificial NNs, have been proposed in the literature \cite{Mingoti2006,Ceriotti2019}. In the present work, we use a Kohonen network, a type of artificial NN based on unsupervised learning, to study local atomic environments. The utilization of NNs has become a standard tool for regression and classification tasks by using both supervised and unsupervised learning.  Since the introduction of artificial NNs in 1943 to model signal processing in the brain \cite{McCulloch1943}, they have been used in many fields, including Biology, Physics, Engineering, and Economics \cite{Rabunal2006}. Applications of NN techniques to domains in physics cover fields such as Particle Physics, Cosmology, Quantum Computing, and Materials Science \cite{Carleo2019}. The advantage of Kohonen networks is that they can be easily implemented, provide a more robust learning and are less sensitive to noise than the k-means algorithm.

In Section \ref{sec:methods}, we introduce the concept of NN and how they are trained. The construction of the NN for cluster analysis is reviewed and discussed in this section, and it is applied to different materials and structures in Section \ref{sec:results}. 

\section{Kohonen networks for cluster analysis}\label{sec:methods} 
\subsection{Kohonen network}
\begin{figure}[h!]
\centering
\includegraphics[width=1\columnwidth]{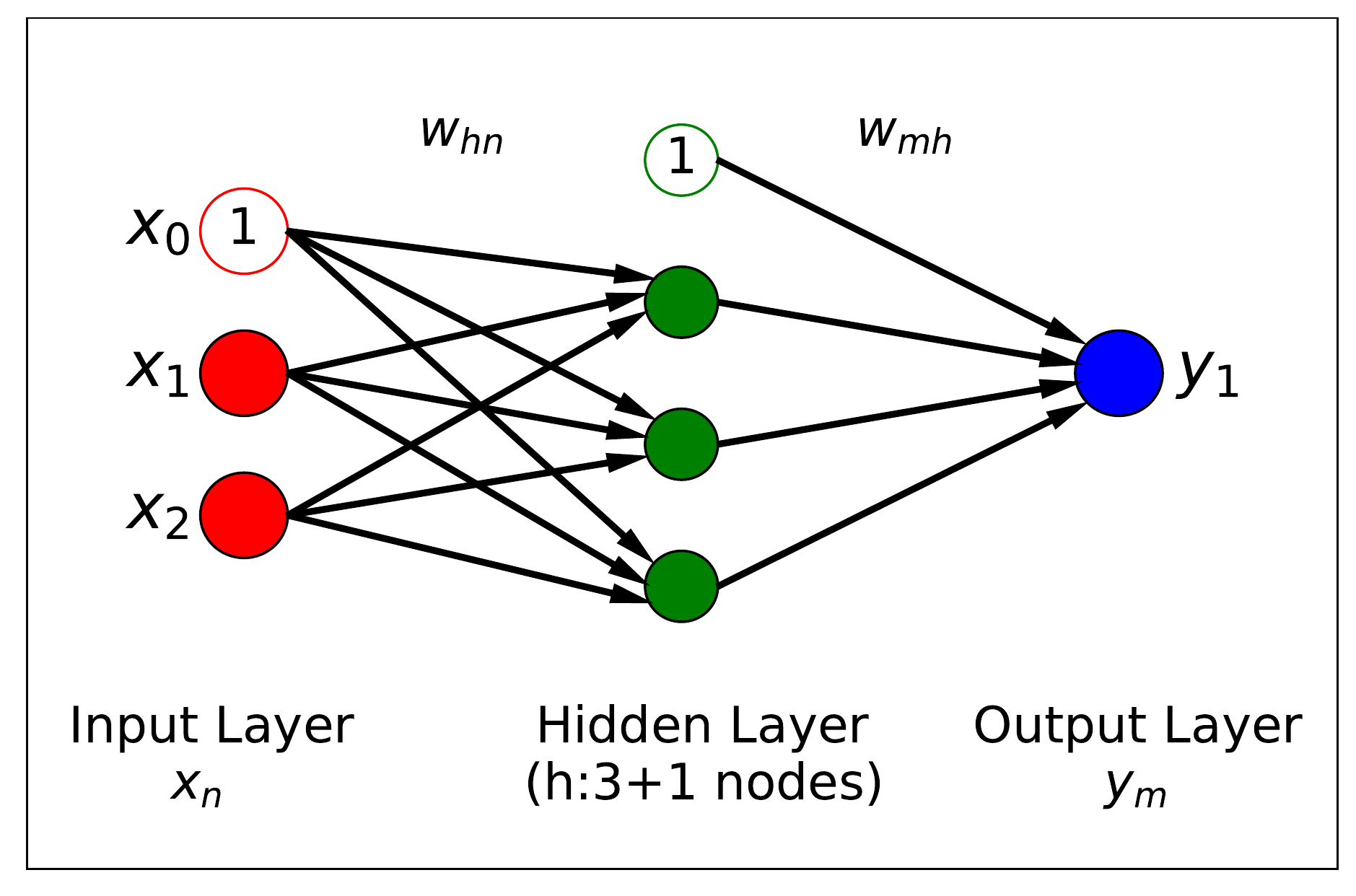}
\caption{ Neural network formed by 2 input features and 1 output feature and a hidden layer formed by 3 nodes. The nodes with white background represent biases. The matrices $w_{hn}$ and $w_{mh}$ connect the layers.} 
\label{fig:ffnn}
\end{figure}

An artificial NN is a collection of connected nodes, known as neurons, which reproduce the interactions among neurons in a biological brain. A simple feed-forward NN is represented in Fig. \ref{fig:ffnn}. In a feed-forward NN, the data flow from input nodes, which contain the input data, to the output nodes through layers of interconnected nodes (if any), known as hidden layers. In the example in Fig. \ref{fig:ffnn}, there are $n=2$ input features, $h=3$ nodes at the hidden layer and  $m=1$ output feature. The circles with white background represent bias nodes, which produce a constant input ($=1$). They allow obtaining more flexibility in the NN. When the information moves forward, the output of each neuron is computed by some non-linear function, $f$, known as activation function, of the weighted sum of its inputs. Activation functions determine the final output of the NN and help normalize the output of each neuron NNs, limiting the presence of large weights. The most common activation functions are the sigmoid function, the hyperbolic function, the ReLu function, and the identity function \cite{FausettLaurene1994}. The NN represented in Fig. \ref{fig:ffnn} corresponds to the analytic form:
\begin{equation}
y_m=f\left(\sum_{h=0}^3w_{mh}f\left(\sum_{n=0}^2w_{hn}x_n\right)\right),
\end{equation}
where $w_{mh}$ and $w_{hn}$ are the weight matrices that connect the layers. NNs are trained using the training data to determine these weights.  The advantage of the architecture of a feed-forward NN is that these weights can be easily obtained using the Back-Propagation algorithm in supervised learning \cite{Larose2014}. In this algorithm, weights are determined by propagating the errors backward, from the output layer to the input layer. 

ClasSOMfier implements a Kohonen network. A Kohonen network is a type of self-organizing map (SOM) that was introduced in 1982 by Teuvo Kohonen for image and sound analysis \cite{Kohonen1982}. SOMs are artificial NNs trained using unsupervised learning \cite{Larose2014}. In this method, input vectors are mapped into a discrete number of output nodes, and the algorithm is based on competitive learning: all nodes compete among themselves to be activated and the winning node determines the category of the input vector.  The objective of this method is not to predict the output vector, but to determine the output node that best represents the input data. The number of nodes is indicated as an input parameter in ClasSOMfier and depends on the degree of variability that the user wants to detect. In Section \ref{sec:results}, we will analyze the performance of this code with a number of clusters between $2$ and $5$, and we will see that these numbers of nodes are enough to detect lattice defects. SOMs are feed-forward NNs but with no hidden layers, i.e., the data flow from the input layer directly to the output layer, and the output layer is represented by a 1- or 2-dimensional lattice of nodes. In this work, we use a 1-dimensional lattice like that shown in Fig. \ref{fig:som}. In this figure, the input layer is represented by the vector $x_n$, where $n=1,2,3$ is the number of input features and is connected with the output layer through the weight matrix $w_{mn}$, where $m=1,2,3,4$ is the number of output nodes, which determines the number of classes. In this method, all weights compete among themselves to determine the winning output feature. This algorithm can be summarized as follows:

\begin{figure}[h!]
\centering
\includegraphics[width=0.75\columnwidth]{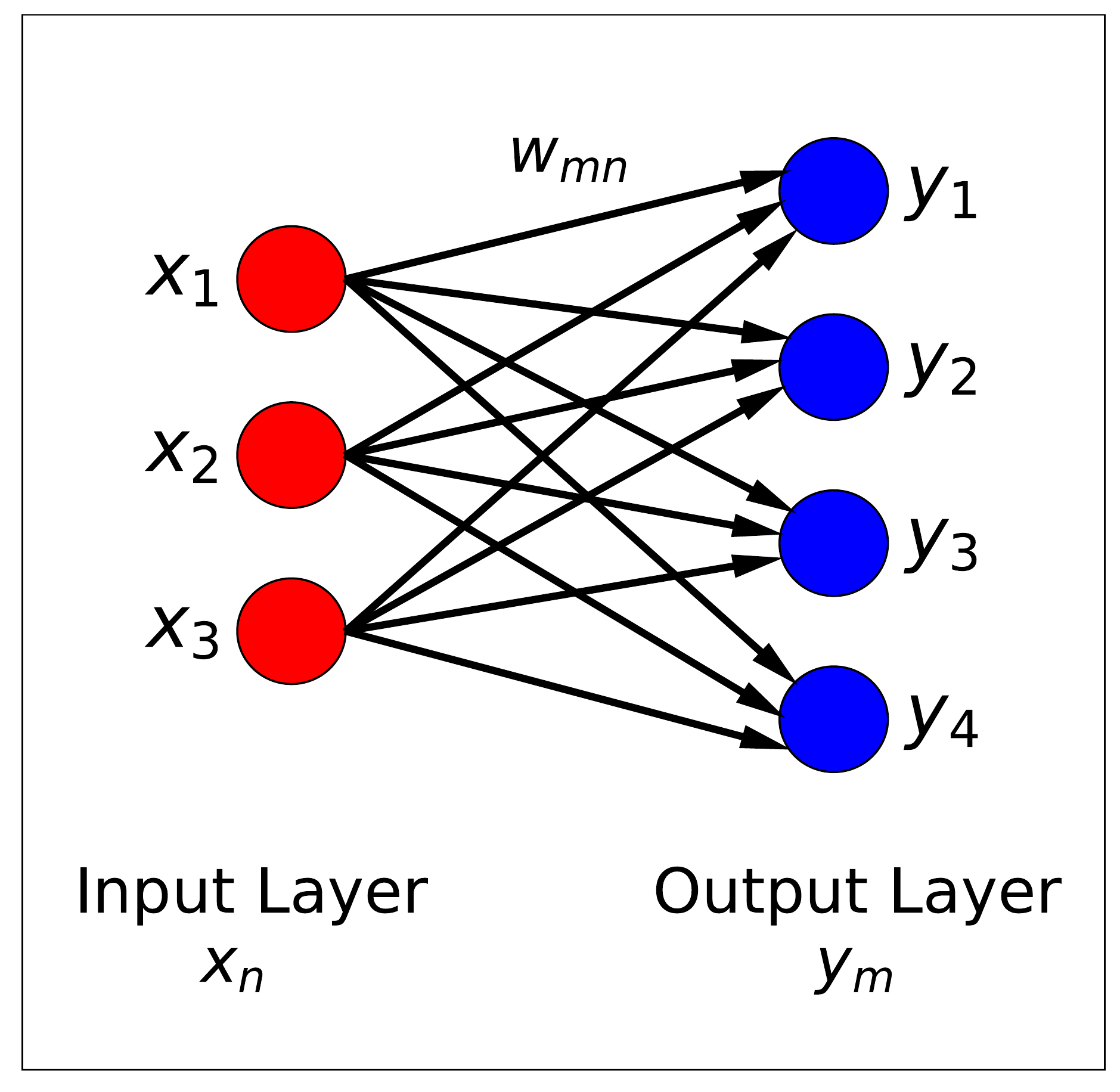}
\caption{ Kohonen network formed by 3 input features and 4 output nodes. The input and output layers are connected through the weight matrix $w_{mn}$.} 
\label{fig:som}
\end{figure}

\begin{enumerate}
\item Initialize all weights $w_{mn}$ to a random value.
\item Set $t=1$, where $t$ is the number of the current iteration. $t_\textnormal{max}$ is the maximum number of iterations (epochs) over all input vectors.
\item Select an input vector $\vec{x}$.
\item For each output node $m$, calculate the Euclidean distance:
\begin{equation}
D_m=\sqrt{\sum_i^n (x_i-w_{mi})^2}.
\label{eq:euclideandistanceerror}
\end{equation}
\item Select the node $m$ that minimizes this distance. This node is known as the Best Matching Unit (BMU). The BMU will determine the category of the input vector.
\item In Kohonen learning, all weights are updated following the rule:
\begin{equation}
w_{mn}^{t}=w_{mn}^{t-1}+\eta(t)h_m(t)\left(x_n-w_{mn}^{t-1}\right),
\end{equation}
where $\eta(t)$ is the learning rate. Different alternatives have been proposed to ensure convergence and prevent oscillations \cite{Riese2020}. In this work, the learning rate decreases with the number of iterations, $t$, as follows:
\begin{equation}
\eta(t)=\eta_0\frac{1}{t}
\end{equation}
Different options have also been tested, with success:
\begin{equation}
\eta(t)=\eta_0e^{-t/t_\textnormal{max}},
\end{equation}
\begin{equation}
\eta(t)=\eta_0^{t/t_\textnormal{max}}.
\end{equation}
These options are compared in Fig. \ref{fig:learningrate}, where $\eta_0$ is the maximum learning rate. 

The function $h_m(t)$ is known as the neighborhood distance weight and determines the variation of the weight after each step. In a Kohonen network, this parameter is higher for all nodes closer to the BMU. In this work, we use the Pseudo-Gaussian neighborhood distance weight proposed by Matsushita et al. \cite{Matsushita2010}:
\begin{equation}
h_m(t)=\exp\left(-\frac{d^2}{2\sigma(t)^2}\right),
\end{equation}
where $d$ is the euclidean distance between node $m$ and the BMU, using periodic boundary conditions (PBC), and $\sigma(t)$ is the neighborhood function, which depends on the current iteration $t$ as the learning rate does. 
\begin{figure}[h!]
\centering
\includegraphics[width=0.75\columnwidth]{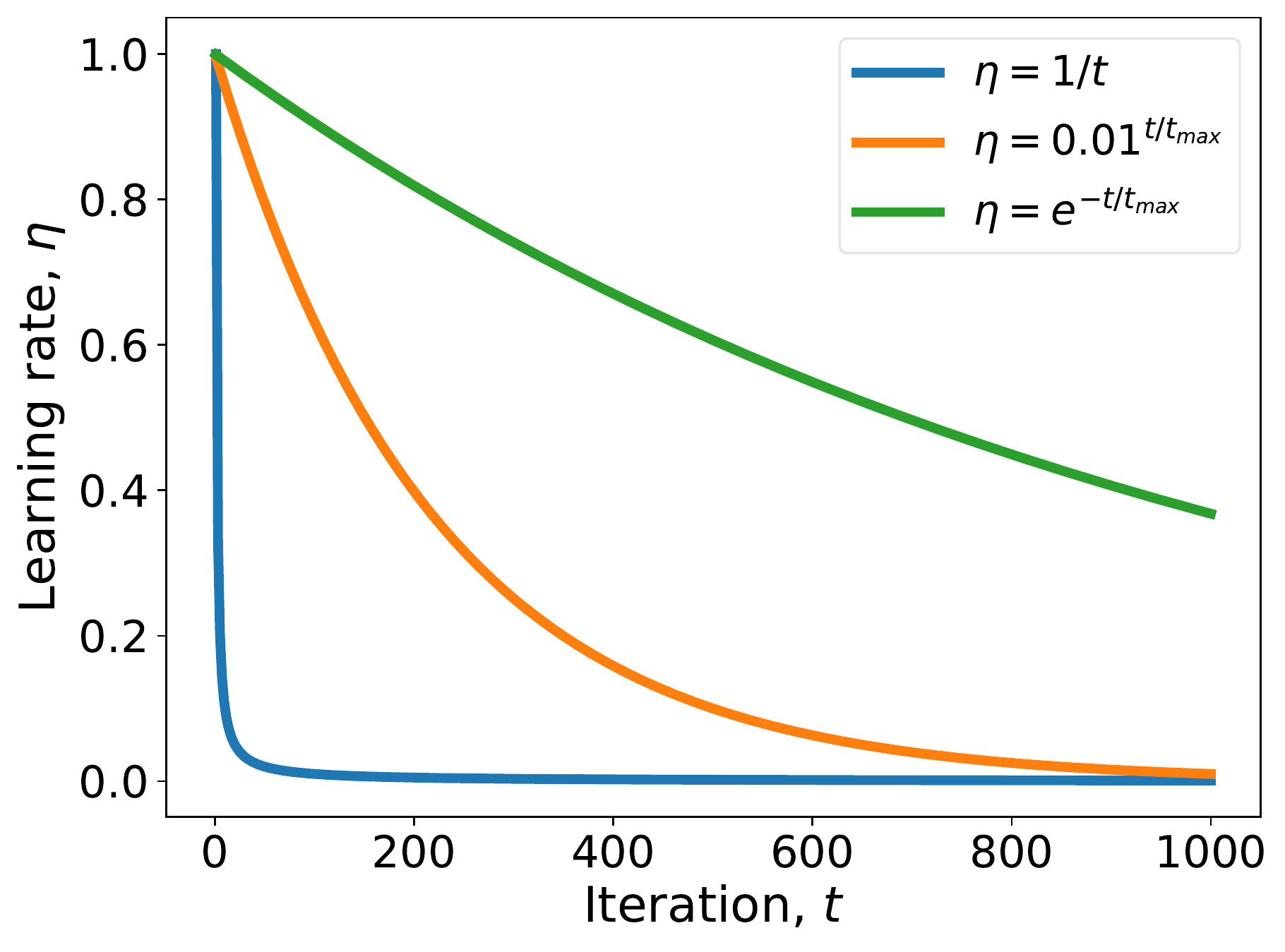}
\caption{Dependence of the learning rate, $\eta$, on the iteration step $t$, where $t_\textnormal{max}$ is the maximum number of iterations.} 
\label{fig:learningrate}
\end{figure}
\item Repeat steps (iii)-(vi) with different input vectors from the train data.
\item Stop if the termination criteria is met or update $t=t+1$ and move to step (iii).
\end{enumerate}

\subsection{Structural description: input vectors}
ClasSOMfier is a software package for cluster analysis. Thus, known the positions of one atom and its neighbors, ClasSOMfier can classify it according to a common pattern shared with all atoms in the same group. The input vector must contain all the relevant information of the atomic environment. Several descriptors of local environments have been proposed in the literature, including Coulomb matrices \cite{Rupp2012}, sine matrices \cite{Faber2015a}, Atom-centered Symmetry Functions \cite{Behler2011} and Smooth Overlap of Atomic Orbitals (SOAP) \cite{Bartok2013}. These descriptors are invariant to rotations, translations and permutations of atoms of the same type, and provide a reliable description of local environments. In the present work, we use the energy per atom and Atom-centered Symmetry Functions to describe local environments. These Symmetry Functions have been proved to be effective in the description of atomic environments in the development of NN potentials to describe atomic interactions  \cite{Behler2011}. Two radial symmetry functions are used to describe the local environment of each atom $i$ \cite{Behler2011}:
\begin{equation}
G_i^1=\sum_j^Nf_c(R_{ij}),
\label{eq:g1}
\end{equation} 
\begin{equation}
G_i^2=\sum_j^Ne^{-\alpha (R_{ij}-R_s)^2}f_c(R_{ij}),
\label{eq:g2}
\end{equation}
where
\begin{equation}
f_c(R_{ij})=0.5\left(\cos\left(\frac{\pi R_{ij}}{R_c}\right)+1\right),
\end{equation}
and $N$ is the number of neighbors inside the sphere with cutoff radius $R_c$ and centered at atom $i$. The function  $f_c(R_{ij})$ is known as cutoff function and is multiplied by Gaussians with width proportional to $\alpha$ and center shifted by a distance $R_s$ in Eq. \ref{eq:g2}. Equations 
\ref{eq:g1} and \ref{eq:g2} are radial functions, and $R_c$, $R_s$ and $\alpha$ are model parameters. The input vector in our calculations consists of a set of $30$ radial symmetry functions, $G_i^1$ and $G_i^2$, with $R_c=0.55l_c,0.75l_c,1.1l_c,1.4l_c, 1.6l_c$ and $1.9l_c$, where $l_c$ is a characteristic length, $R_s=0.75R_c$ and $\alpha=1,0.5, 0.1,0.01$ \AA$^{-1}$, plus the energy per atom and the 6 numbers of neighbors in the 6 spheres with radii $R_c$. The number of input features necessary to describe the local environment of one atom depends on the complexity of the material.

\section{Results and Discussion}\label{sec:results} 
Python is the default interfacing language through which the user interacts with the algorithm, which is implemented in Fortran. The code reads dump files from LAMMPS  \cite{lammpsWeb,Plimpton1995} and returns xyz files that can be visualized in Ovito \cite{ovito2010}.  In this section, we use ClasSOMfier to detect lattice defects in PbTe and Mg. The structures are generated in LAMMPS after performing energy minimization or MD simulations with the potentials described in Refs. \cite{javpabjor2019} and \cite{Dickel2018} for PbTe and Mg, respectively. An implementation of the classifier to separate atoms into two different groups according to their local environment is shown in Fig. \ref{fig:python}. $\sigma$, $\eta$ and the number of epochs can be added as optional arguments. In this example, default values are: $\sigma=1.0$ \AA, $\eta=0.5$ and $500$ epochs. The computational time depends on the system size and number of nodes. In this example, the computational time is 5 s. The source code is directly available on Github
at \url{https://github.com/JaviFdezT/ClasSOMfier}.

\begin{figure}[h!]
\begin{lstlisting}[language=Python,frame=single]
from classomfier import ClasSOMfier

#Initializes the classifier with lc=6.44 and 2 groups. The thirs argument is the input file.
nn=ClasSOMfier(6.44,2,"dumpgb.file")  
#Classifies all atoms into two groups. The output file can be found at "./data/output.xyz" (default location).
nn.execute()
\end{lstlisting}
\label{fig:python}
\caption{Python example to run the classifier with default values. $l_c$ is the characteristic length. The input file, dumpgb.file, includes the positions and energies of all atoms in a structure containing grain boundaries. The total number of atoms is 6447.} 
\end{figure}

The performance of the ClasSOMfier package is illustrated for the detection of lattice defects in PbTe and Mg. PbTe presents a rocksalt structure, with cubic symmetry, while Mg presents an hcp structure.  In this section, we will study the effect of charged particles in PbTe and characterize complex grain boundaries in Mg.

\subsection{Study of Point Defects}
\begin{figure}[h!]
\centering
\subfigure[]{%
\includegraphics[width=0.33\columnwidth]{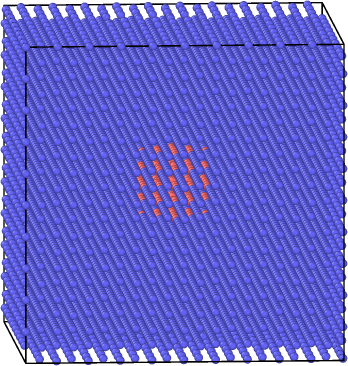}\hspace{2cm}
\label{fig:vsinglepbte1}}
\quad
\subfigure[]{%
\includegraphics[width=0.33\columnwidth]{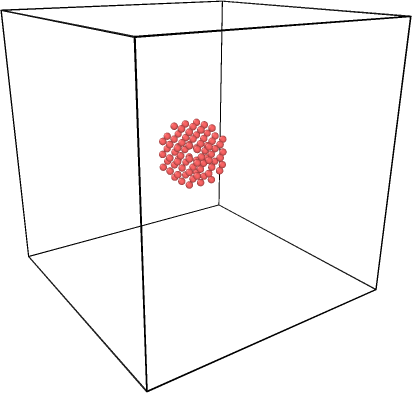}
\label{fig:vsinglepbte2}}
\subfigure[]{%
\includegraphics[width=0.33\columnwidth]{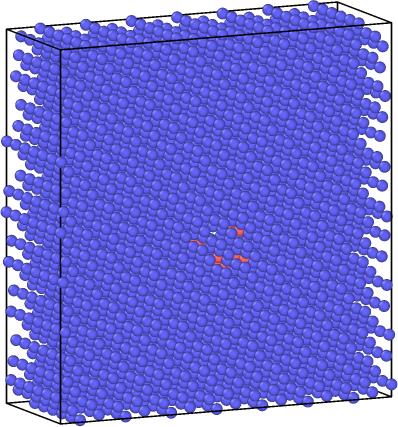}\hspace{2cm}
\label{fig:vsinglemg1}}
\quad
\subfigure[]{%
\includegraphics[width=0.33\columnwidth]{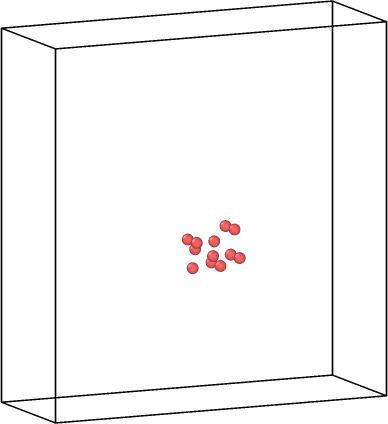}
\label{fig:vsinglemg2}}
\caption{Study of the area of influence of a single vacancy in PbTe (a,b) and Mg (c,d). The vacancy is created at the center of a simulation box containing 8000 atoms, and all atoms are classified into two groups. Atoms in blue present the ideal lattice structure and atoms in red are affected by the presence of the vacancy. The red regions are isolated in (b,d).} 
\label{fig:singlevacancy}
\end{figure}

We demonstrate the use of the classifier to detect point defects and study their region of influence. The input data is generated with LAMMPS. In Fig. \ref{fig:singlevacancy}, we analyze the effect of a single vacancy in PbTe and Mg. We can see that the presence of a single vacancy has a stronger impact in PbTe than in Mg. This is caused by the fact that the total charge is not kept constant in PbTe. In Fig. \ref{fig:vsinglepbte1}, one atom of Pb has been deleted at the center of the simulation box, and a group of atoms around the vacancy is affected by this new defect (red atoms). On the other hand, the blue atoms keep their rocksalt structure and their local environment is different from that of the atoms in red. As a consequence, the two types of atoms are classified into two different groups. In Fig. \ref{fig:vsinglepbte2}, the blue atoms have been deleted to visualize the region of influence of the vacancy. As we can see in Figs. \ref{fig:vsinglepbte2} and  \ref{fig:vsinglemg2}, the number of atoms affected by the presence of a vacancy is smaller in Mg, because a single vacancy in PbTe is a charged defect. It is important to note that the detection of lattice defects is not affected by the use of the energy per atom, which depends on the atom type, as a feature in the input vector.

\begin{figure}[h!]
\centering
\subfigure[]{%
\includegraphics[width=0.3\columnwidth]{figure5b.png}
\label{fig:vsinglepbte3}}
\subfigure[]{%
\includegraphics[width=0.3\columnwidth]{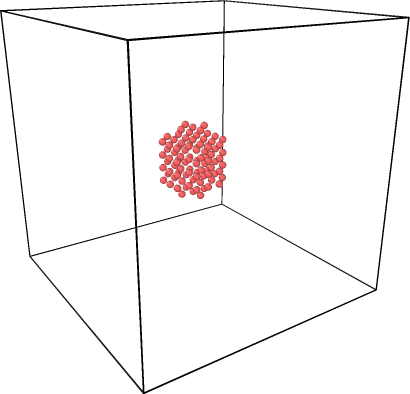}
\label{fig:vdimer}}
\subfigure[]{%
\includegraphics[width=0.3\columnwidth]{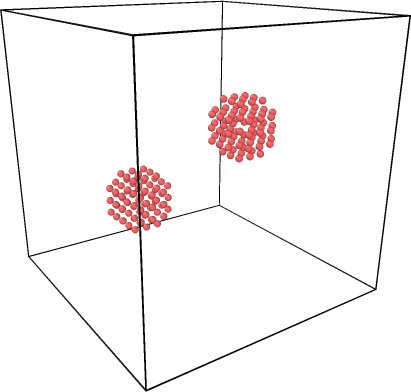}
\label{fig:vpair}}
\caption{Study of the area of influence of a single vacancy (a), a Schottky dimer (b), and a Schottky pair (c) in PbTe. The vacancies are created at the center of a simulation box containing 8000 atoms, and all atoms are classified into two groups. Only the atoms affected by the presence of the point defects are shown.} 
\label{fig:divpbte}
\end{figure}

In Fig. \ref{fig:divpbte}, we study  the effect of the charge on the formation of vacancies. We compare the region of influence of a single vacancy (Fig. \ref{fig:vsinglepbte3}), a Schottky dimer (Fig. \ref{fig:vdimer}), and a Schottky pair (Fig. \ref{fig:vpair}) when the atoms forming the ideal lattice structure are not shown. A Schottky pair is formed by a pair of isolated vacancies, and the Schottky dimer is formed by 2 consecutive vacancies. The region of influence of the Schottky dimer is formed by 92 atoms, while it is formed by 160 atoms for a Schottky pair. This difference is in good agreement with the fact that the formation energy to form a Schottky pair is larger than that to form a Schottky dimer.

\begin{figure}[h!]
\centering
\subfigure[]{%
\includegraphics[width=0.3\columnwidth]{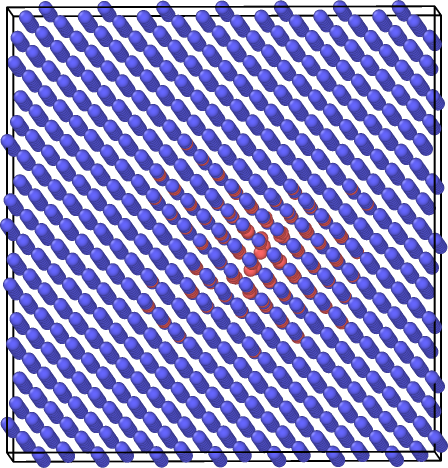}
\label{fig:interstmg1}}
\subfigure[]{%
\includegraphics[width=0.3\columnwidth]{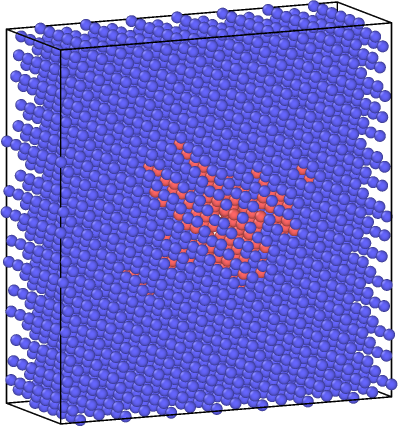}
\label{fig:interstmg2}}
\subfigure[]{%
\includegraphics[width=0.3\columnwidth]{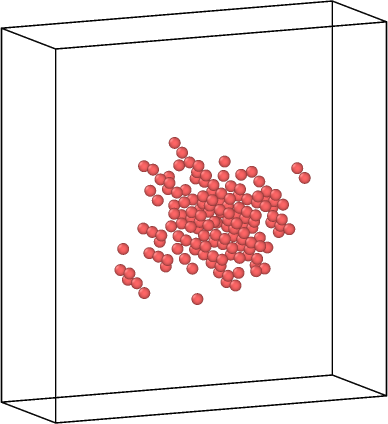}
\label{fig:interstmg3}}
\caption{Study of the area of influence of a single interstitial in Mg. The interstitial is created at the center of a simulation box containing 9600 atoms, and all atoms are classified into two groups. Atoms in blue present the ideal lattice structure and atoms in red are affected by the presence of the vacancy. The red region is isolated in (c).} 
\label{fig:interstmg}
\end{figure}

\begin{figure}[h!]
\centering
\subfigure[]{%
\includegraphics[width=0.3\columnwidth]{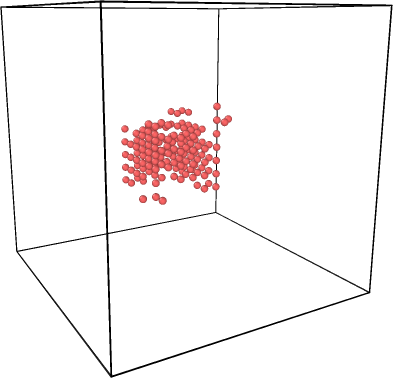}
\label{fig:interstpbte1}}
\subfigure[]{%
\includegraphics[width=0.3\columnwidth]{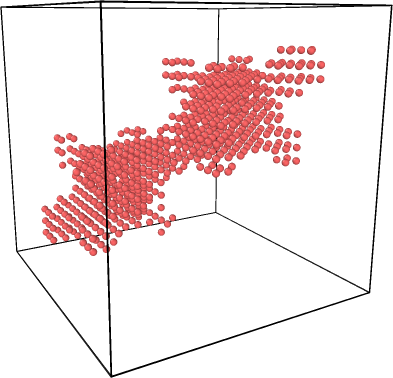}
\label{fig:interstpbte2}}
\subfigure[]{%
\includegraphics[width=0.3\columnwidth]{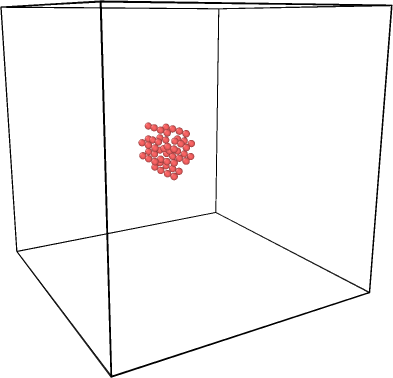}
\label{fig:interstpbte3}}
\caption{Study of the area of influence of a single interstitial (a), a pair of interstitials (b), and a dimer of interstitials (c) in PbTe. The interstitials are created at the center of a simulation box containing 8000 atoms, and all atoms are classified into two groups. Only the atoms affected by the presence of the point defects are shown.} 
\label{fig:interstpbte}
\end{figure}

We also demonstrate the use of ClasSOMfier to detect interstitials and study their region of influence. We analyze the effect of a single interstitial in Mg in Fig. \ref{fig:interstmg}, and a single interstitial and pairs of interstitials in PbTe in Fig. \ref{fig:interstpbte}. In Figs. \ref{fig:interstmg1} and \ref{fig:interstmg2}, a single interstitial is created at the center of the simulation box containing 9600 atoms. In Fig. \ref{fig:interstpbte}, we compare the effect of a single vacancy (Fig. \ref{fig:interstpbte1}), a pair of interstitials (Fig. \ref{fig:interstpbte2}) and a dimer of interstitials (Fig. \ref{fig:interstpbte3}). We can see that the presence of a single interstitial has a stronger impact in Mg than in PbTe; this difference occurs because Mg presents an hcp lattice, which is not isotropic. We also observe that fewer atoms are affected by dimers than by single interstitials. This difference lies in the fact that dimers are no charged defects, and therefore long-range Coulomb interactions are not present. 

\subsection{Study of Volume Defects}
\begin{figure}[h!]
\centering
\subfigure[]{%
\includegraphics[width=0.25\columnwidth]{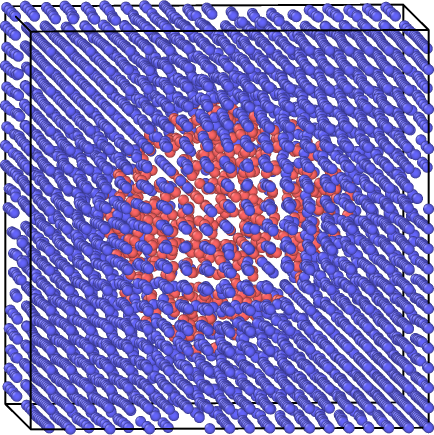}
\label{fig:voidpbte1}}
\subfigure[]{%
\includegraphics[width=0.42\columnwidth]{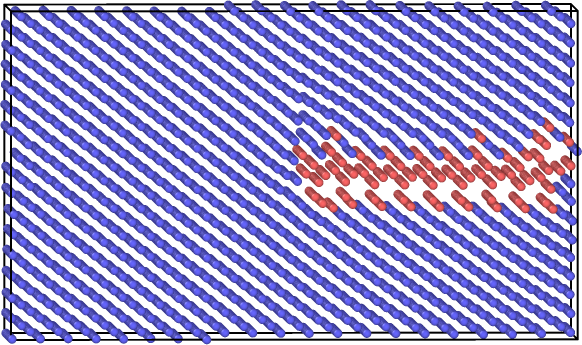}
\label{fig:dislocationpbte}}
\subfigure[]{%
\includegraphics[width=0.26\columnwidth]{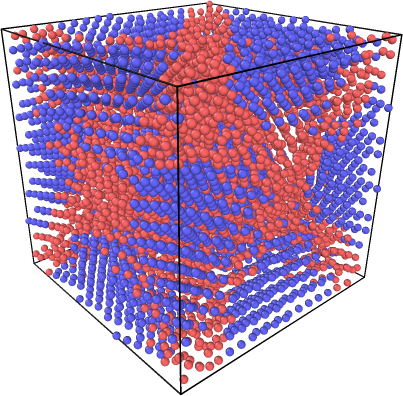}
\label{fig:gbpbte}}
\caption{Study of the area of influence of a void (a), a dislocation (b) and grain boundaries (c) in PbTe. The void is created at the center of a simulation box containing 8000 atoms, followed by relaxation using MD, and all atoms are classified into two groups. Atoms in blue present the ideal lattice structure and atoms in red are affected by the presence of lattice defects.} 
\label{fig:volumepbte}
\end{figure}

We used ClasSOMfier to analyze the region of influence of a void, dislocations, and grain boundaries in PbTe in Fig. \ref{fig:volumepbte}. In Fig. \ref{fig:voidpbte1}, a void is created at the center of a simulation box containing 8000 atoms, followed by relaxation using MD with the classical potential defined in Ref. \cite{javpabjor2019}. ClasSOMfier identifies a thin layer around the void (in red) as the regions formed by the atoms affected by the presence of the void. In Fig. \ref{fig:dislocationpbte}, ClasSOMfier is used to characterize the region of influence of a dislocation in PbTe. The edge dislocation was generated with Atomsk \cite{Hirel2015} and relaxed with LAMMPS.  The code can detect and differentiate small deviations in the ideal lattice structure, associated with the presence of the dislocation. In Fig. \ref{fig:gbpbte}, ClasSOMfier differentiates atoms with the ideal lattice structure (blue) and atoms at grain boundaries (red). The polycrystalline structure is formed by randomly generated grains following the Voronoi tessellation \cite{Aurenhammer1991}. The study and characterization of grain boundaries at the nanoscale is important to understand phenomena such as grain growth and Zener pinning \cite{F.JHumphreys2004}. Grain boundaries are metastable defects and, at finite temperatures, grains grow until a limiting grain size is reached \cite{F.JHumphreys2004}. The analysis of the interaction between grain boundaries and volume defects present in the sample, together with the study of changes in the grain boundary width and the position of the grain boundary, is crucial to understand properties at the meso- and macroscale. ClasSOMfier can isolate grain boundaries to analyze how they evolve in MD simulations.

\begin{figure}[h!]
\centering
\subfigure[]{%
\includegraphics[width=0.3\columnwidth]{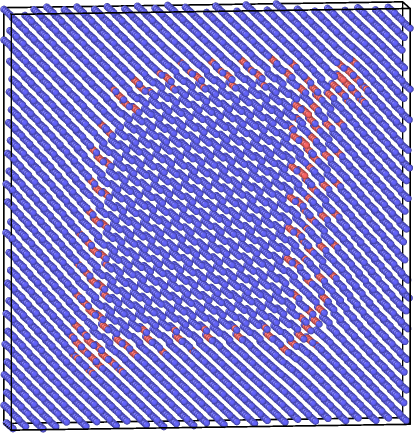}
\label{fig:gbmg1}}
\subfigure[]{%
\includegraphics[width=0.3\columnwidth]{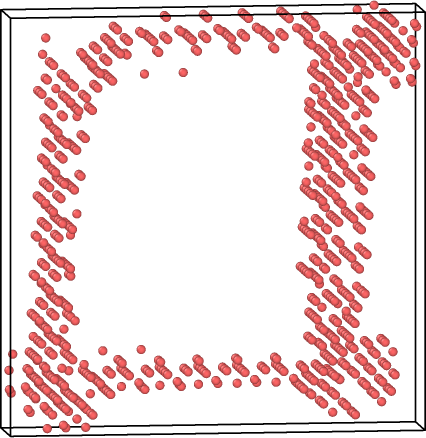}
\label{fig:gbmg2}}
\subfigure[]{%
\includegraphics[width=0.3\columnwidth]{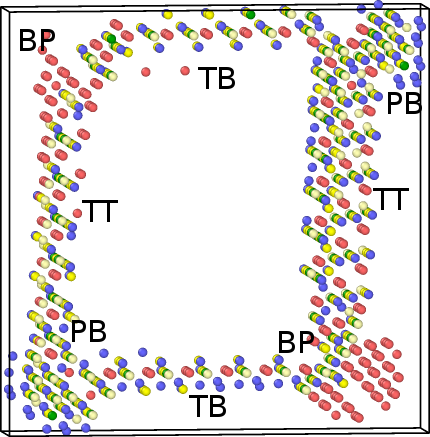}
\label{fig:gbmg3}}
\caption{Study of the area of influence of a grain boundary in Mg. The atoms are classified into several groups according to their local environment. The atoms forming the grain boundary, in red, are isolated in (b), while the atoms from the other groups are shown in blue in (a). In (c), the atoms at the grain boundary is classified into different groups (5) to detect common patterns} 
\label{fig:volumemg}
\end{figure}

In Fig. \ref{fig:volumemg}, ClasSOMfier is used to analyze a more complex structure. A twin embryo is created in Mg, and this software package is used to detect and characterize the grain boundary. The structure was generated using the Eshelby method \cite{Xu2013,Hu2020} and relaxed using the potential described in Ref. \cite{Dickel2018}. The final structure is shown in Fig. \ref{fig:gbmg1}. The atoms forming the ideal lattice (blue) and the grain boundary (red) are shown in Fig. \ref{fig:gbmg1}, and the atoms at the grain boundary are isolated in Fig. \ref{fig:gbmg2}. Finally, all atoms at the grain boundary are analyzed in  Fig. \ref{fig:gbmg3}. There are different types of grain boundaries between the two grains due to the different relative orientations between the lattice vectors of the adjacent grains at each interface. The different types of grain boundaries are shown in  Fig. \ref{fig:gbmg3}: two twin boundaries (TBs) and two twin tips (TTs) forming the lateral grain boundaries, and two basal-prismatic (BP) grain boundaries and two prismatic-basal (PB) grain boundaries at the corners. The program is not able to differentiate these regions; however, it is possible to detect specific patterns for each grain boundary. In this figure, the atoms at the grain boundary are classified into five groups, shown in red, yellow, white, blue and green. It is possible to see that the red atoms are dominant at BP and non-existent at PB, while blue atoms are dominant at PB and non-existent at BP. Additionally,  TB boundaries are thin and present red atoms at the center, while TT boundaries are thicker and red atoms are not at the center.

\section{Conclusions}\label{sec:conclusions}
ClasSOMfier is a new code developed in Python and Fortran for cluster analysis and detection of lattice defects. Its user-friendly interface allows using it in Python with a comprehensive workflow, from LAMMPS to Ovito. It reads dump files from LAMMPS containing atom positions and energies per atom, classifies all atoms into a given number of categories, and writes output files that can be read by Ovito to visualize the different clusters. It implements a Kohonen network, which is a NN based on unsupervised learning. The utilization of NNs has become a standard tool for regression tasks in materials science \cite{Behler2016}, and we show that it is a powerful tool for cluster analysis and detection of lattice defects. Its robust architecture can easily find patterns in lattice structures and the figures shown in Section \ref{sec:results} do not depend on the initial conditions, i.e., the initial weights. This problem is often found in k-mean clustering. 

Radial functions, coordination numbers and the energy per atom are used to build the input vector. These radial functions have been proved to be reliable structural descriptors \cite{Behler2011}. The performance of these radial functions and the Kohonen network implemented in ClasSOMfier was studied through the analysis of the effect of point and volume defects in Mg, which presents an hcp lattice that can give way to complex grain boundaries, and PbTe, a binary compound formed by charged atoms. In these simulations, the number of clusters is set to 2 to differentiate the perfect lattice from regions affected by the presence of lattice defects. This number can be increased if more patterns are present in the structure, depending on the number of patterns that the user wants to distinguish. If the number of nodes is large enough, discontinuous clusters associated with the atom type or the distance to the equilibrium positions can be found. However, this is not the main objective of the code, and the number of outputs is set as an input parameter.

ClasSOMfier can detect lattice defects and their region of influence. These lattice defects include vacancies, interstitials, voids, dislocations and grain boundaries. The presence of lattice defects affects the positions and energies of all neighbor atoms, and ClasSOMfier detects them by differentiating common patterns in the system.
ClasSOMfier can also detect common patterns in amorphous structures. 

\ack
This work was supported by a research grant from Science Foundation Ireland (SFI) and the Department for the Economy Northern Ireland under the SFI-DfE Investigators Programme Partnership, Grant Number 15/IA/3160. We thank Jorge Kohanoff, Vladyslav Turlo, Yang Hu and Eduardo M. Bringa for insightful discussions.

\section*{References}
\providecommand{\newblock}{}


\end{document}